# scientific reports

**OPEN**

# Altered structural balance of resting-state networks in autism

Z. Moradimanesh[1], R. Khosrowabadi[1], M. Eshaghi Gordji[1,2] & G. R. Jafari[1,3]✉

What makes a network complex, in addition to its size, is the interconnected interactions between elements, disruption of which inevitably results in dysfunction. Likewise, the brain networks' complexity arises from interactions beyond pair connections, as it is simplistic to assume that in complex networks state of a link is independently determined only according to its two constituting nodes. This is particularly of note in genetically complex brain impairments, such as the autism spectrum disorder (ASD), which has a surprising heterogeneity in manifestations with no clear-cut neuropathology. Accordingly, structural balance theory (SBT) affirms that in real-world signed networks, a link is remarkably influenced by each of its two nodes' interactions with the third node within a triadic interrelationship. Thus, it is plausible to ask whether ASD is associated with altered structural balance resulting from atypical triadic interactions. In other words, it is the abnormal interplay of positive and negative interactions that matters in ASD, besides and beyond hypo (hyper) pair connectivity. To address this question, we explore triadic interactions based on SBT in the weighted signed resting-state functional magnetic resonance imaging networks of participants with ASD relative to healthy controls (CON). We demonstrate that balanced triads are overrepresented in the ASD and CON networks while unbalanced triads are underrepresented, providing first-time empirical evidence for the strong notion of structural balance on the brain networks. We further analyze the frequency and energy distributions of different triads and suggest an alternative description for the reduced functional integration and segregation in the ASD brain networks. Moreover, results reveal that the scale of change in the whole-brain networks' energy is more narrow in the ASD networks during development. Last but not least, we observe that energy of the salience network and the default mode network are lower in ASD, which may be a reflection of the difficulty in dynamic switching and flexible behaviors. Altogether, these results provide insight into the atypical structural balance of the ASD brain (sub) networks. It also highlights the potential value of SBT as a new perspective in functional connectivity studies, especially in the case of neurodevelopmental disorders.

Human brain is an inherently complex network even at rest, comprising nearly 10 billion neurons connected by about 100 trillion synapses[1]. Yet, the brain owes its beautiful complexity and the consequent cognitive capacities not solely to the size but also the interconnected interactions between its constituent elements. In the last two decades, graph theory has provided a valuable framework to study the structure of the complex brain networks[2]. The C. elegance connectome was modeled as a binary graph, as one of the first contact points between neuroscience and modern network science[3]. The tract-tracing data of cat and macaque monkey was examined in early 2000[4] and later graph theory started to be well applied to the human neuroimaging data[5]. Along with the progress in neuroimaging techniques, there has also been growing interest in investigating associations between time series extracted from the brain regions, i.e., functional connectivity. Altogether, graph-theoretic studies are an important foundation for computational modeling of the complex brain networks, and have revealed fundamental properties of their organization and function, in addition to their alterations in brain disorders[6,7]. While all these advances and the potential of graph theory's perspective is undisputed, a vital question have yet to be asked: Suppose in a signed brain network $x$, $y$ and $z$ regions are connected, what is the inevitable impact of $xz$ and $yz$ interactions on the sign and weight of the interaction between $x$ and $y$? Is it realistic to study the $xy$ interaction independent from the triadic interconnection ($xyz$) it lives in? Specially in case of heterogeneous brain disorders such as ASD, a complex neurodevelopmental disorder with substantial heterogeneity not only in multiple causes and courses but also in the onset[8–11]. Recently, this concern has been well addressed by the long-standing structural balance theory (SBT) in variety of scenarios, and triadic interactions are accepted to play key role in organization of real-world signed networks[12–16]. Here, we apply concepts from SBT to the weighted signed rs-fMRI networks of individuals with ASD compared to healthy controls in order to answer the following questions: Considering recent verification of SBT by many real-world signed networks, do the brain networks display structural balance as well? Regarding controversies on strong and weak notions of SBT, which will be confirmed on the brain networks? Knowing the previously proposed hypo (hyper) functional connectivity in ASD, if we take into account the interplay between positive and negative links, which types of triadic interactions (in terms of both the frequency and energy distribution) are atypical in ASD? Is the consequent structural balance altered for the whole brain ASD network in course of development? What about the functional sub-networks? These questions in essence, are the cornerstone of this study.

[1]Institute for Cognitive and Brain Sciences, Shahid Beheshti University, Tehran, Iran. [2]Department of Mathematics, Semnan University, 35195-363, Semnan, Iran. [3]Department of Physics, Shahid Beheshti University, Tehran, Iran. ✉email: g_jafari@sbu.ac.ir

Initially proposed by Heider[17] and mathematically formulated by Cartwright and Harary[18], SBT has been a promising approach in arguing why the structures of many real signed social[12–14], political[15], ecological[16] and biological[19] networks are the way they are. The distinguishing feature of SBT is that it highlights the role that balanced (unbalanced) triadic interactions play in forming the global structure of a network. This theory has led to better understanding of how a tendency to reduce overall stress directs and arranges the organization of signed networks. Specifically, SBT argues that a link between two nodes being positive (negative) is hugely affected by a third node's presence within a triadic structure. Accordingly, four types of triads are defined in this context (Eq. 1, Fig. 6), namely, strongly balanced $T_3$ : $(+ + +)$, weakly balanced $T_1$ : $(+ - -)$, strongly unbalanced $T_2$ : $(+ + -)$, and weakly unbalanced $T_0$ : $(- - -)$, details of which is provided in methods. Frequent building blocks of two to five vertices, known as the structural and functional motifs, have been identified and investigated in the brain networks through applying framework's other than SBT as well[20,21]. However, there are two main differences between triads as motifs from one hand and triads as has been defined in SBT from the other hand: (1) Sign of links are not of concern in motifs, yet they are at the heart of triads' definition in the context of SBT, (2) Motifs are mostly directed structures while triads in SBT are not originally. Moreover, measures of segregation such as the clustering coefficient or transitivity are also based on the frequency of triads, yet again they measure properties of a network regardless of signs[22]. This is while, it has been recently confirmed on the real-world networks that studying the interplay between positive and negative links within triadic interactions, instead of focusing exclusively on positive links, opens new doors to existing challenges[12,13].

Signed networks not only represent the structure but also embody additional information on the state of relationships between nodes, and has been long of interest to network scientists in different domains from social science to biology. In the last decade, many social scientists have reported that in analyzing large-scale social networks considering the content of interactions, positive (negative) links representing friendship (enmity), contains much promise. In this regard, the key role that triadic interactions play in forming the global structure of signed social networks have been strongly confirmed[23–25]. Similarly in biological and biochemical signed networks the positive (negative) ties correspond to activating, correlating (inhibiting, anti-correlating) interactions[26]. There has been various evidences of anti-correlated functional sub-networks in the intrinsic brain networks of both human[27] and animal[28] as well. These studies imply that negative associations are meaningful in the resting-state brain networks and not only the unwelcome result of preprocessing steps, such as the global signal regression[29]. Although it has been shown that regressing the global signal introduces anti-correlations to the network[30], thus here we did not regressed the global signal to make sure existing negative links are in a way reflections of neural activities. To explore real-world signed networks, SBT as both a general theory as well as a practical framework for conducting experiments has been very reassuring. Since its proposal, there has been two main directions in the literature of SBT: (1) Theoretically extending its analytical aspects[31–35], (2) Empirically verifying it on real signed networks[13,16], or both[12,14]. Moreover, a recent study has investigated the brain evidence for SBT from psychological viewpoint[36], this is while we investigate the network evidence for the theory through examining Blood Oxygen Level-Dependent (BOLD) signals. Furthermore, besides SBT which investigates the structural balance of complex networks through studying triadic interactions, there has been valuable research to explore higher-order interactions in the brain networks based on topological properties as well[37–39].

In this study, we explore the weighted signed rs-fMRI networks of individuals with ASD compared to healthy participants in the context of SBT. We conduct our analysis in three age ranges, namely, 1st childhood (6–9 years old), 2nd childhood (9–13 years old) and adolescence (13–18 years old). First, according to previous empirical studies, *we expect the brain networks to display structural balance as well*—**hypothesis #1**, meaning that balanced triads are overrepresented while unbalanced triads are underrepresented compare to chance. Although, due to controversies regarding the over (under) representation of $T_0$, it is nontrivial whether the strong or weak notion of balance would be confirmed. Next, we study the frequency of pair and triadic interactions in the brain networks. Considering previous results on hypo (hyper) functional connectivity in the ASD brain network, it seems that *taking into account sign of links and exploring triadic interactions provides us with a more contextual insight into atypical functional connectivity in the ASD brain network*—**hypothesis #2**. Moreover, we illustrate the energy distributions of triads and suppose that *there are probably of note differences regarding energy distributions of triads between ASD and CON networks*—**hypothesis #3**. However, it is not clear which types of triads are different in terms of energy and during which age ranges. Last but not least, we examine the overall energy of the whole-brain network and 17 Yeo sub-networks and expect to observe *alterations in the energy of the ASD brain (sub) networks, possibly suffering from lower or more limited energy scales in some of the key (sub) networks*—**hypothesis #4**. Our study, while appreciating standard approaches in analyzing functional connectivity, proposes a new perspective in exploring rs-fMRI brain networks, which can be specially of interest in investigating complex brain disorders such as ASD.

| | 1st Childhood | | | | 2nd Childhood | | | | Adolescence | | | |
|---|---|---|---|---|---|---|---|---|---|---|---|---|
| | $|T_i|$ | $p(T_i)$ | $p_0(T_i)$ | $s(T_i)$ | $|T_i|$ | $p(T_i)$ | $p_0(T_i)$ | $s(T_i)$ | $|T_i|$ | $p(T_i)$ | $p_0(T_i)$ | $s(T_i)$ |
| **(A) Number of balanced and unbalanced triads in the brain networks of CON group** | | | | | | | | | | | | |
| $T_3 : (+ + +)$ | 525,438.06 | 0.628 | 0.527 | 188.41 | 1,214,646.87 | 0.471 | 0.389 | 271.72 | 1,622,249.62 | 0.446 | 0.375 | 286.13 |
| $T_1 : (+ - -)$ | 104,304.64 | 0.124 | 0.094 | 96.51 | 537,459.93 | 0.208 | 0.164 | 194.93 | 796,741.18 | 0.219 | 0.172 | 236.52 |
| $T_2 : (+ + -)$ | 200,416.79 | 0.239 | 0.369 | − 246.39 | 781,750.31 | 0.303 | 0.423 | − 392.92 | 1,143,696.25 | 0.315 | 0.427 | − 433.69 |
| $T_0 : (- - -)$ | 5493.50 | 0.006 | 0.008 | − 17.12 | 44,986.90 | 0.017 | 0.022 | − 49.43 | 68,595.85 | 0.018 | 0.024 | − 66.21 |
| **(B) Number of balanced and unbalanced triads in the brain networks of ASD group** | | | | | | | | | | | | |
| $T_3 : (+ + +)$ | 498,880.87 | 0.651 | 0.563 | 158.12 | 1,417,970.87 | 0.479 | 0.406 | 261.37 | 1,491,940.00 | 0.483 | 0.405 | 282.66 |
| $T_1 : (+ - -)$ | 76,009.00 | 0.099 | 0.080 | 62.01 | 571,349.62 | 0.193 | 0.157 | 173.51 | 588,859.37 | 0.190 | 0.154 | 176.36 |
| $T_2 : (+ + -)$ | 186,927.28 | 0.244 | 0.349 | − 194.79 | 918,819.62 | 0.310 | 0.415 | − 365.79 | 955,916.50 | 0.309 | 0.419 | − 390.14 |
| $T_0 : (- - -)$ | 4086.83 | 0.005 | 0.006 | − 11.84 | 51,197.86 | 0.017 | 0.021 | − 47.66 | 49,637.08 | 0.016 | 0.020 | − 49.16 |

**Table 1.** Number and probability of triads in the brain networks compared to the null model. On average, in both CON and ASD groups, both types of balanced triads, $T_3$ and $T_1$, are overrepresented (positive $s(T_i)$), while both types of unbalanced triads, $T_2$ and $T_0$, are underrepresented (negative $s(T_i)$). $|T_i|$, total number of $T_i$; $p(T_i)$, fraction of $T_i$ in the brain network; $p_0(T_i)$, fraction of $T_i$ in the null model; $s(T_i)$, the amount of surprise, i.e., is the number of standard deviations by which the actual number of $T_i$ differs from its expected number under the null model; CON, control; ASD, autism spectrum disorder.

## Results

**Empirical evidence of SBT on the brain networks.** First, we have investigated the notion of structural balance in the brain networks. According to the strong version of this notion, real-world networks evolve in a way that eventually unbalanced triads are underrepresented while balanced triads are overrepresented. The weak notion however, is less strict and argues that $T_0$ triads, although unbalanced yet may be overrepresented as well. To this aim, we have applied a method proposed by Leskovec et al.[12] which states that for a triad $T_i$ if $p(T_i) > p_0(T_i)$ then $T_i$ is overrepresented, and if $p(T_i) < p_0(T_i)$ then $T_i$ is underrepresented. Here $p_0(T_i)$ is the fraction of triads of type $T_i$ after shuffling, details of which are provided in methods. Furthermore, to investigate how big these are (over) under representations are, we have applied the concept of surprise, i.e, $s(T_i)$, which on the order of tens would be significant, due to the distribution of $s(T_i)$ being standard normal.

As our results show, on average in the brain networks of both CON (Table 1A) and ASD (Table 1B) groups and in all three age ranges both balanced triads are overrepresented relative to chance, that is for $T_3$ and $T_1$ we have $p(T_i) > p_0(T_i)$. On the contrary, on average both unbalanced triads are underrepresented compared to the null model, that is, we have observed $p(T_i) < p_0(T_i)$ for both $T_2$ and $T_0$ triads. Additionally, all the surprises have been significant. It should be mentioned that these results are group-level, that is, averaged over all participants in each group. Results on the level of each participant is the same for $T_3$, $T_2$ and $T_1$ triads, meaning that for each and every participant in all groups we have overrepresentation regarding $T_3$ and $T_1$ while underrepresentation for $T_2$ triads. However, for $T_0$ triads a small percentage of participant's networks have more $T_0$ triads relative to chance as follows: for ASD and CON groups during 1st childhood the percentage of networks with overrepresentation of $T_0$ triads were 22.2% and 10%, respectively. During 2nd childhood this percentages were 3.8% in ASD group and 5.7% in CON group. Finally, during adolescence in ASD group 3.4% and in CON group only 1.7% of networks had more $T_0$ triads relative to chance.

**Why considering triadic interactions in studying the brain networks?** After verifying SBT on the brain networks, we have explored the value of examining ternary interactions (triads) in the brain network along with the pair interactions (links). In other words, why consider triadic interactions while analyzing brain networks? To study this question, we have first compared mean differences in the frequency of positive and negative links between ASD and CON groups in three age ranges. Then, we have conducted the same analysis for the mean frequency of different types of triads. As results show, although exploring links and triads are both valuable in revealing statistically significant differences between groups, acknowledging triads provides results with bigger effect sizes, that is, practical significance, details of which are as follows.

First, to compare the mean differences in the frequency of links between ASD and CON groups during development, we have conducted two two-way ANCOVAs, one with the frequency of positive and the other with negative links as a dependent variable (Table 2A). We have considered group and age as independent variables while controlling for FIQ, medication, mean frame-wise displacement as head motion parameter and site information. We are interested in the main effect of group and the interaction between group and age. While the former provides an overall difference between ASD and CON groups, the latter allows us to explore the differences during development. Then, we have conducted the same analysis for triads, that is, we have defined four two-way ANCOVAs one for each type of triads as has been defined in Eq. 1 (Table 2B). Another option was to conduct two MANCOVAs, one for links and the other for triads, however due to multicollinearity between the two types of links and the four types of triads we chose ANCOVA over MANCOVA. As results show (Table 2), the main effect of group is neither significant on the mean frequency of links nor triads. However, while the effect of interaction between group and age is only statistically significant for positive links (a medium effect), it is significant for all types of triads. More interestingly, it is only practically significant on the mean frequency of $T_2$ triads ($F(1, 251) = 95.17$ $p < 0.001$, $partial\eta^2 = 0.46$).

| Source | Dependent variable | $p$ val | Partial $\eta^2$ |
|---|---|---|---|
| **(A) ANCOVA regarding the frequency of links** | | | |
| Group | Positive links | 0.18 | 0.008 |
| | Negative links | 0.21 | 0.007 |
| Group * Age | Positive links | < 0.001* | 0.13 |
| | Negative links | 0.08 | 0.02 |
| **(B) ANCOVA regarding the frequency of triads** | | | |
| Group | $T_3 : (+ + +)$ | 0.55 | 0.002 |
| | $T_1 : (+ - -)$ | 0.05 | 0.01 |
| | $T_2 : (+ + -)$ | 0.56 | 0.001 |
| | $T_0 : (- - -)$ | 0.69 | 0.001 |
| Group * Age | $T_3 : (+ + +)$ | < 0.001** | 0.11 |
| | $T_1 : (+ - -)$ | < 0.001** | 0.08 |
| | $T_2 : (+ + -)$ | < 0.001** | 0.46 |
| | $T_0 : (- - -)$ | < 0.03* | 0.03 |

**Table 2.** Analysis of covariance for the frequency of links versus triads. (A) The effect of group is not significant on the mean frequency of links. The effect of interaction between group and age is only significant for the positive links, yet not practically significant. (B) For triads regarding the effect of interaction between group and age, not only all the differences are statistically significant but the effect size (partial $\eta^2$) is interestingly considerable for $T_2$. **$p < 0.01$, *$< 0.05$ (Bonferroni corrected).

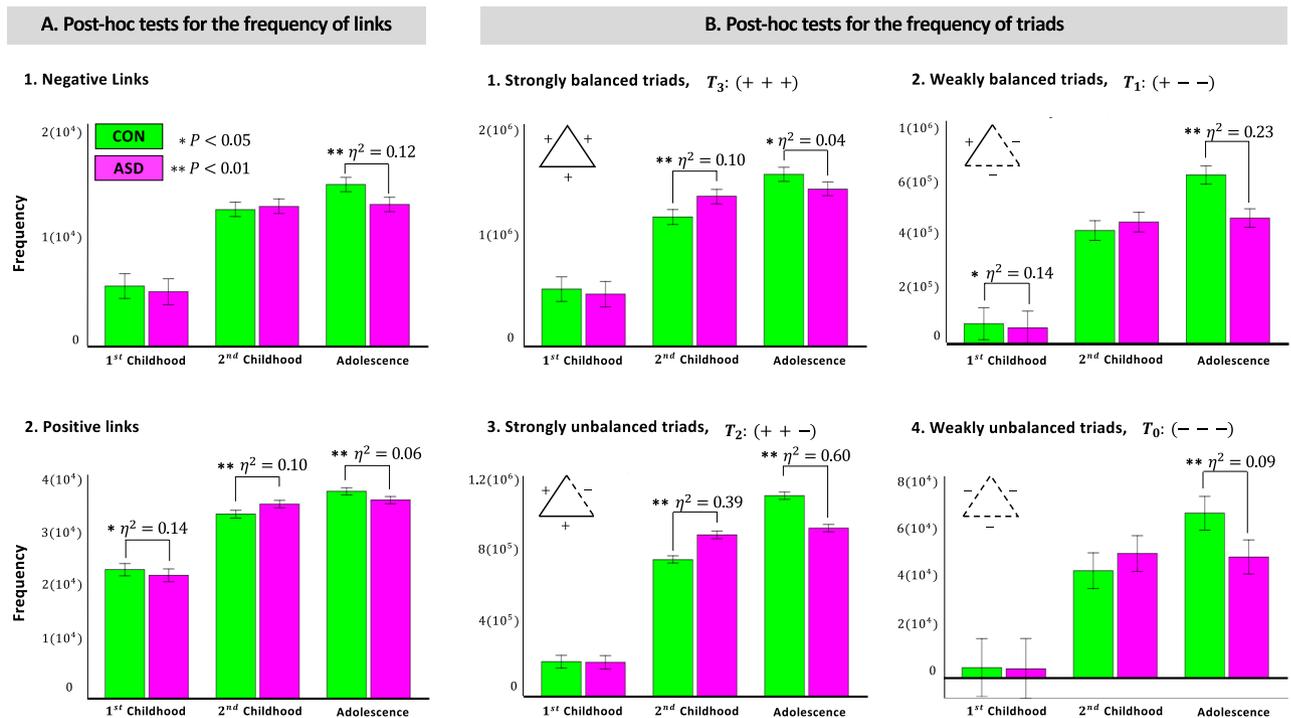

**Figure 1.** Mann-Whitney U test on the frequency of links and triads. (**A**) Difference between ASD and CON groups is only practically significant ($\eta^2 \geq 0.14$) for the mean frequency of positive links during 1st childhood. (**B**) Yet, considering triads provides us with practical differences not only during 1st childhood but also during 2nd childhood and adolescence. $\eta^2$, the effect size; CON, control; ASD, autism spectrum disorder (Color Online).

To further explore in which age levels the differences in mean frequency of links (Fig. 1A) and triads (Fig. 1B) have occurred between ASD and CON groups, we have conducted post-hoc tests using Mann-Whitney U test. As results depict, during 1st childhood the frequency of positive links from one hand ($U(20, 18) = 102.00$, $z = -2.28$, $p = 0.02$), and $T_1$ triads from the other hand ($U(20, 18) = 101.00$, $z = -2.31$, $p = 0.02$) are both practically significant ($\eta^2 = 0.14$). During 2nd childhood, while there is a medium difference ($\eta^2 = 0.10$) in the mean frequency of positive links ($U(52, 52) = 1853.00$, $z = 3.25$, $p = 0.001$), there is a large difference ($\eta^2 = 0.39$) in the mean frequency of $T_2$ triads ($U(52, 52) = 2338.00$, $z = 6.41$, $p < 0.001$). Additionally, there is also a medium difference ($\eta^2 = 0.10$) in the mean frequency of $T_3$ triads between ASD and CON groups during 2nd childhood ($U(52, 52) = 1868.00$, $z = 3.35$, $p = 0.001$). Similarly, for adolescents although there is a significant difference for both types of links (for positive links: $U(58, 58) = 1181.50$, $z = -2.76$, $p = 0.006$ and for negative links: $U(58, 58) = 1010.00$, $z = -3.71$, $p < 0.001$), yet both of these differences are medium, $\eta^2 = 0.06$ and $\eta^2 = 0.12$ for positive and negative links, respectively. However, surprisingly for $T_2$ triads the difference is far more bigger than both types of links ($U(58, 58) = 162.00$, $z = -8.39$, $p < 0.001$, $\eta^2 = 0.60$). There is also a big difference ($\eta^2 = 0.23$) in the mean frequency of $T_1$ triads ($U(58, 58) = 745.00$, $z = -5.17$, $p < 0.001$). Additionally, there is a medium difference ($\eta^2 = 0.09$) in the mean frequency of $T_0$ triads ($U(58, 58) = 1102.00$, $z = -3.20$, $p = 0.001$), and a small difference ($\eta^2 = 0.04$) in the mean frequency of $T_3$ triads ($U(58, 58) = 1270.00$, $z = -2.27$, $p = 0.02$).

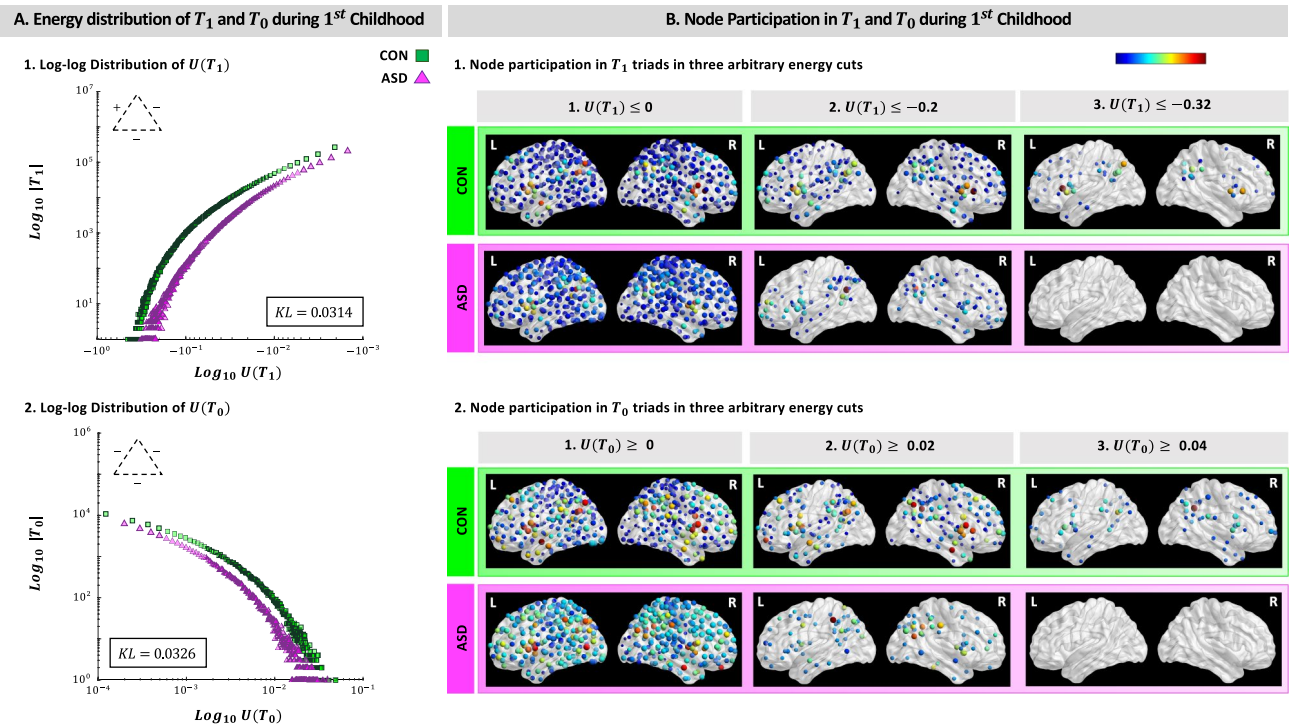

**Figure 2.** Energy distributions and node participation in $T_1$ and $T_0$ triads during 1st childhood. (**A**) The energy distributions of ASD group (purple/dark triangles) lag behind CON group (green/light squares) for $T_1$ (A.1) and $T_0$ (A.2) triads during 1st childhood. (**B**) For both triads we have observed a threshold above which node participation is zero for the brain network of ASD group, yet nonzero for CON group (B.1.3 for $T_1$ and B.2.3 for $T_0$). A detailed version of B.1.3 and B.2.3 with regions of interest's labels is provided in the Supplementary Fig. S2. BrainNet Viewer 1.63[40] (https://www.nitrc.org/projects/bnv) has been used for visualization of the brain networks. $|T_i|$, total number of $T_i$; $U(T_i)$, the energy of $T_i$; KL, the Kullback–Leiber divergence; CON, control; ASD, autism spectrum disorder (Color Online).

**Energy distribution of triads.** After studying the frequency of triads in the brain networks of ASD and CON groups, we have explored the energy distributions of different types of triads. To this aim, we have calculated the energy of triads, $U(T_i)$, as shown in Fig. 6, and then for each group in three age ranges outlined the corresponding distributions. As results indicate (Supplementary Fig. S1), for all types of triads the brain networks of ASD and CON groups have many triads with small energies and a few with considerable energies. Furthermore, results of comparisons between energy distributions of triads in ASD and CON groups show that during 1st childhood the pattern of energy distributions differs between the two groups for $T_1$ and $T_0$ triads (Fig. 2A). That is, looking at the energy distributions of these two triads in ASD group (purple/dark triangles) compare to CON group (green/light squares) it is clear that the energy distributions in ASD group lag behind distributions as they are in CON group ($KL = 0.03$, for both types). For other types of triads and during 2nd childhood and adolescence we have not observed such a lag (Supplementary Fig. S1).

To better clarify this lag, for each $T_i$ we have investigated node participation which for any given node specifies: (1) In how many $T_i$ this node has participated? and (2) How big are the energies of those $T_i$s? In Fig. 2B and Supplementary Fig. S2 we have addressed the former question with size of a given node and the latter with its color, which is a color-map from blue to red that respectively corresponds to the minimum and maximum levels of triads' energy. For example, a big blue node is present in many $T_i$s which are small in terms of energy. On the other hand, a small red node although lives in just a few $T_i$s but in those that are important in terms of energy levels. As can be seen, moving from the minimum to maximum level of energy (Fig. 2B.1,B.2; left to right), while node participation decreases sharply to zero in ASD group, the speed of this decrease is slower in CON group. In other words, we have observed a threshold in the energy distributions of both $T_1$ and $T_0$ triads above which node participation is zero for ASD group yet nonzero for CON group. To be specific, in ASD group no regions of interest participate in creating $T_0$ triads that have energies higher than 0.04, while in CON group many nodes are participating so that the resulting network has $T_0$ triads with energies higher than 0.04. This pattern holds for the case of $T_1$ triads. That is, while in ASD group node participation is zero for $T_1$ triads with | *Energy* | $\geq 0.32$, the network of CON group during 1st childhood enjoys the presence of $T_1$ triads with | *Energy* | $\geq 0.32$ (Supplementary Fig. S2).

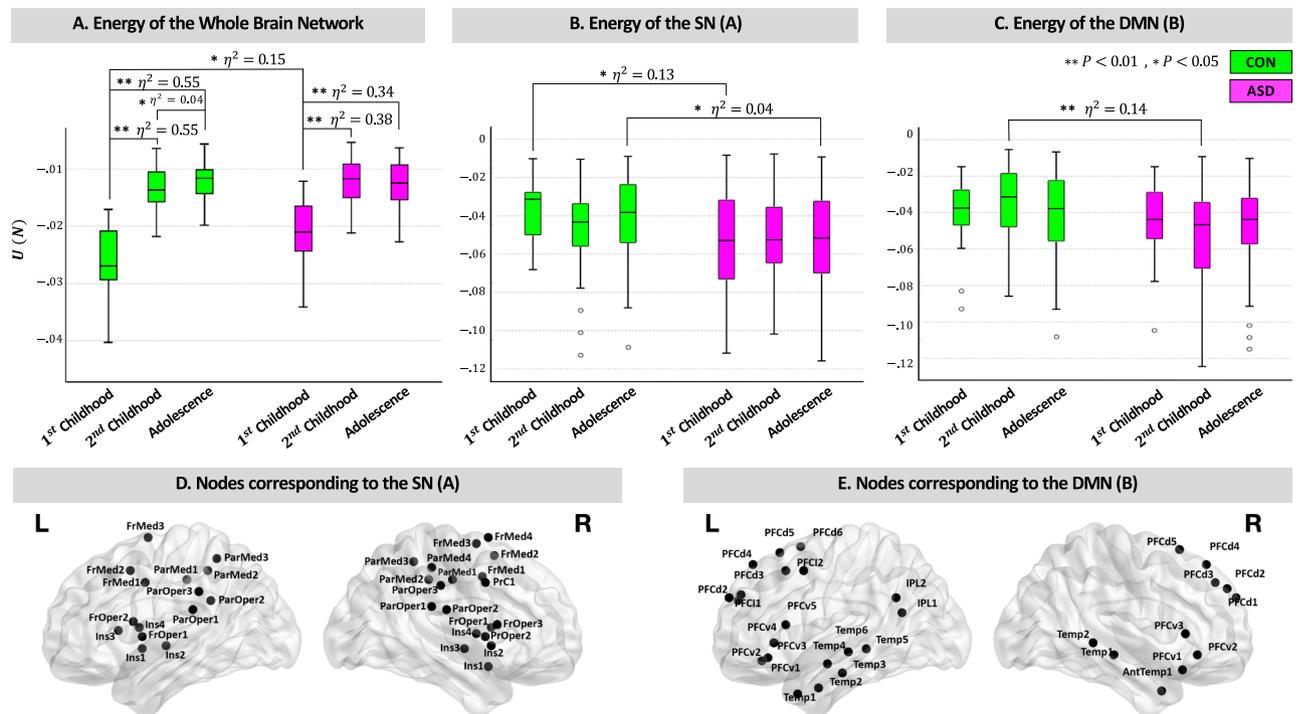

**Figure 3.** Energy of the brain (sub) networks during development. (**A**) For both groups, energy of the whole brain network increases monotonically with age, yet ANCOVA results show that the pattern of this increase is different between the two groups. (**B,C**) The effect of age on energy of the SN (**A**) and DMN (**B**) is significant, with ASD having less energies in both sub-networks. Results of Mann-Whitney U tests with corresponding effect sizes ($\eta^2$) are shown above each of the box plots. (**D,E**) Regions of interest related to the SN (**A**) and DMN (**B**), as has been defined in the Schaefer atlas. BrainNet Viewer 1.63[40] (https://www.nitrc.org/projects/bnv) has been used for visualization of the brain networks. SN, the salience/ventral attention network; DMN, the default mode network; Ins, the insula; PFC, the prefrontal cortex; Temp, the temporal cortex; FrOper, the frontal opercular; ParOper, the parietal opercular; FrMed, the frontal medial; IPL, the inferior parietal lobule; CON, control; ASD, autism spectrum disorder (Color Online).

### Structural balance of the whole brain network and 17 Yeo sub-networks.
Next, we have analyzed energy levels of the brain networks in the context of SBT. In this regard, first we have calculated the energy levels of each participant's network using Eq. 2. For each individual, we have computed 18 different energies, one for the whole brain network and the rest 17 each corresponding to 17 Yeo sub-networks. Then, we have explored the effect of group and its interaction with age on these energy levels controlling for FIQ, medication, mean frame-wise displacement as head motion parameter and site information. We have performed this by conducting 18 ANCOVAs each regarding one of the energy levels as mentioned above, results of which are as follows:

- The effect of group on the whole brain energy is statistically significant ($F(1, 251) = 12.08$, $p < 0.001$, $partial \eta^2 = 0.05$). Similarly, the effect of interaction between group and age on the whole brain energy is significant as well ($F(2, 251) = 4.57$, $p = 0.01$, $\eta^2 = 0.04$).
- The effect of group is significant on the energy levels of the SN (**A**) ($F(1, 251) = 12.52$, $p < 0.001$, $partial \eta^2 = 0.05$) and the DMN (**B**) ($F(1, 251) = 6.73$, $p = 0.01$, $partial \eta^2 = 0.03$).

Moreover, to identify the groups to which these differences are related to we have conducted post-hoc tests using Mann-Whitney U test. Results for the energy levels of the whole brain networks are as follows (Fig. 3A):

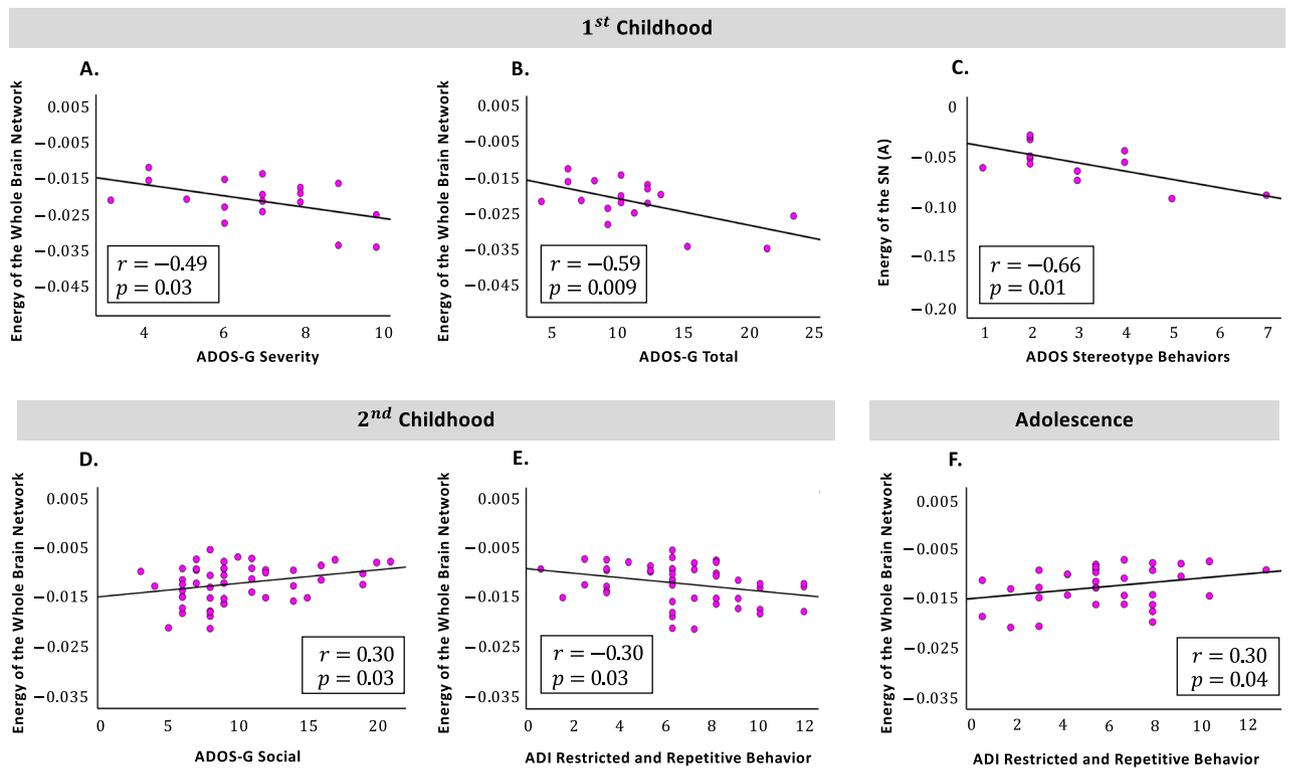

**Figure 4.** Pearson's correlation coefficients between network measures and behavioral scores. (**A**–**C**) During 1st childhood, ADOS-G severity and total scores show negative associations with the whole brain energy. As like, energy of the SN (**A**) is negatively associated with ADOS stereotype behaviors. (**D**,**E**). During 2nd childhood, energy of the whole brain network positively correlates with ADOS-G social scores and negatively with ADI restricted behaviors. (**F**) For adolescents energy of the whole brain network is positively correlated with ADI restricted behaviors. SN, the salience/ventral attention network; ADI, the autism diagnostic interview; ADOS, the autism diagnostic observation schedule; ADOS-G, the autism diagnostic observation schedule-generic.

- During 1st childhood, energy of the whole brain network is significantly greater for ASD than CON group ($U(20, 18) = 261.00, z = 2.36, p = 0.01$) and it is not only statistically but also practically significant ($\eta^2 = 0.15$).
- For both the ASD and CON brain networks, there is a practically significant difference between energy levels during 1st childhood from one hand, with 2nd childhood and adolescence from the other hand. Results of comparison between 1st childhood and 2nd childhood for ASD group is $U(18, 52) = 854.00, z = 5.18, p < 0.001, \eta^2 = 0.38$, and for CON group it is, $U(20, 52) = 1018.00, z = 6.26, p < 0.001, \eta^2 = 0.55$. Additionally, between 1st childhood and adolescence results are as follows for ASD and CON groups, respectively: $U(18, 58) = 938.00, z = 5.08, p < 0.001, \eta^2 = 0.34$ and $U(20, 58) = 1152.00, z = 6.54, p < 0.001, \eta^2 = 0.55$.
- More interestingly, there is a significant difference ($\eta^2 = 0.04$) between 2nd childhood and adolescence in CON group ($U(52, 58) = 1876.00, z = 2.20, p = 0.02$), yet for ASD group there is no such a change in energy level of the whole brain network from 2nd childhood to adolescence, that is, energy of the whole brain network remains statistically unchanged after 1st childhood.

For the Yeo sub-networks, during 1st childhood a Mann-Whitney U test have indicated that energy of the SN (A) is greater for CON group than ASD group ($U(20, 18) = 104.00, z = -2.22, p = 0.02, \eta^2 = 0.13$). Later during adolescence, there is also a significant difference on the energy level of the SN (A) between ASD and CON groups ($U(58, 58) = 1280.00, z = -2.22, p = 0.02, \eta^2 = 0.04$; Fig. 3B). Additionally, there is a practically significant difference ($\eta^2 = 0.14$) on the energy level of the DMN (B) between ASD and CON groups during 2nd childhood ($U(52, 52) = 758.00, z = -3.86, p < 0.001$; Fig. 3C).

**The brain-behavior relationship: correlation of energy levels with behavioral scores.** Ultimately, to investigate whether energy of the whole brain network and Yeo sub-networks are associated with the clinical symptoms of ASD we have measured the correlation between the two (Fig. 4). On the x axis, we have scores from modules defined in three well-known instruments for diagnosing and assessing ASD, that is, the autism diagnostic interview (ADI), the autism diagnostic observation schedule (ADOS) and the autism diagnostic observation schedule-generic (ADOS-G). From these tests, score from some of the modules were available and among them those modules have been chosen that had less missing values. On the y axis, we have energy levels of the brain (sub) networks.

As results of two sided Pearson's correlation between behavioral scores and energy levels indicate, during 1st childhood there was a significant negative association between energy of the whole brain network and ADOS-G severity ($r(18) = -0.49, p = 0.03$) (Fig. 4A), this holds for ADOS-G total scores as well ($r(18) = -0.59, p = 0.009$) (Fig. 4B). Moreover, during 1st childhood energy of the SN (A) was negatively correlated with ADOS stereotype behaviors ($r(13) = -0.66, p = 0.01$) (Fig. 4C). Furthermore, during 2nd childhood and adolescence there were significant correlations between energy of the whole brain network and the following behavioral modules:

- There was a positive correlation with ADOS-G social scores ($r(51) = 0.30, p = 0.03$) during 2nd childhood (Fig. 4D), yet for adolescence the correlation with ADOS communication was significantly negative ($r(32) = -0.32, p = 0.04$). It is worth mentioning that the ADOS communication scores for adolescents were from both module-3 and -4. Moreover, the correlations of each module separately with the whole brain energy did not reach significance, possibly due to missing data. Thus, results regarding this correlation are reported in Supplementary Fig. S3.
- There was a negative correlation during 2nd childhood with ADI restricted and repetitive behaviors ($r(49) = -0.30, p = 0.03$) (Fig. 4E), which was a significantly positive association during adolescence ($r(34) = 0.30, p = 0.04$) (Fig. 4F).

## Discussion

The current study has analyzed the weighted signed brain networks of ASD and CON groups in the context of structural balance theory (SBT) during development. Although analyzing pair interactions between brain regions has revealed fundamental network properties in the last decades, yet in exploring a complex system such as the brain questioning the unavoidable impact of the interactions each of the two regions has with a third region within triadic interrelationships seems plausible. In other words, the crucial role that weighted signed triadic interactions play in the organization of real-world complex networks which have been widely accepted, is worth considering. Accordingly, the current study has demonstrated the following results:

*Evidence for the strong formulation of Heider's balance theory in the brain networks* (Table 1). Our findings have provided first-time empirical evidence for Heider's balance theory on the ASD and healthy brain networks during development. To be specific, we have found that *on average for both ASD and CON groups, from 1st childhood all the way to adolescence, balanced triads are heavily overrepresented in the brain networks while unbalanced triads are underrepresented*—**hypothesis #1**. Our result on the group-level interestingly supports the strong notion of structural balance theory, which states that both unbalanced triads, $T_2$ and $T_0$, are underrepresented in real networks. This is while in social networks, the weak formulation of structural balance has been reported, which states that only unbalanced triads of type $T_2$ are underrepresented in real signed networks[12,13]. While unbalanced triads of type $T_0$ are in some cases (such as the Epinions and Wikipedia) overrepresented and in some although underrepresented yet to a lesser degree than the other three types (such as the Slashdot)[12]. These results are consistent with the fact that on the level of each participant, we have found that there is a small percentage of individuals within each group in the networks of which $T_0$ triads are overrepresented. However, these percentages are small. Thus, the overall underrepresentation of $T_0$ triads seems to be the case in the brain networks.

*Lower frequency of $T_1$ during 1st childhood, higher $T_2$ during 2nd childhood and lower $T_1$ and $T_2$ during adolescence in ASD* (Fig. 1). Regarding practical differences ($\eta^2 \geq 0.14$) in the mean frequency of different types of triads during development, we have observed the following results: (1) During 1st childhood, the mean frequency of $T_1$ triads is significantly greater in CON group. (2) During 2nd childhood, the mean frequency of $T_2$ triads is significantly greater in ASD group. (3) During adolescence, the mean frequency of both $T_1$ and $T_2$ triads are greater in CON group. As results depict, the frequency of $T_3$ triads in which all three links are positive, and $T_0$ triads that are consisted of three negative links, are not of much practical concern when studying differences between the brain network of ASD compared to healthy controls. This can pave the way for a candidate explanation that for the ASD network compared to the CON network regarding links/triads frequency, *it is the interplay of negative and positive links that gives rise to the practical differences between ASD and healthy individuals*—**hypothesis #2**, and not solely the hypo (hyper) connectivity of one type of links independent from the other type and detached from the setting they belong to. Furthermore, the frequency of triads mentioned above in ASD group being lower during 1st childhood (mean age 8) and adolescence (mean age 16), yet higher during 2nd childhood (mean age 11) reflects the hypo (hyper) functional connectivity in the backbone of networks which is pretty much in line with previous studies besides many different controversies in this regard[41–43]. These changes in the pairwise functional connectivity are as well consistent with studies that have suggested critical developmental shift during the time of puberty, which is typically between 9 and 13 years[42,44].

*Lower participation of brain regions in $T_1$ and $T_0$ triads with high energies in ASD during 1st childhood* (Fig. 2). As results indicate, the energy distributions of $T_1$ and $T_0$ triads in ASD group lag behind CON group during 1st childhood. That is, we have observed a threshold in the energy distributions of $T_1$ and $T_0$ triads above which node participation is zero for ASD group, yet nonzero for CON group. In other words, the ASD network lacks high energy $T_1$ and $T_0$ triads compared to healthy individuals during 1st childhood. This can be interpreted according to the role of $T_1$ and $T_0$ triads in the organization of networks from the perspective of SBT, which is to provide connected modularity. Specifically, when a network has $T_1$ or $T_0$ triads on its stable state, we expect it to have different groups of nodes (the so-called modules) with negative links connecting them to each other. This may go quite consistently with the theory of functional segregation and integration in the brain networks[2]. According to this theory, functional segregation corresponds to the presence of specialized modules or clusters within the brain network. In comparison, functional integration in the brain is the potential to combine specific information from local distributed brain regions. *The Absence of high energy $T_1$ and $T_0$ triads in the ASD network (compared* *to healthy individuals) may provide us with an alternative explanation for the reduced functional integration and segregation in ASD*—hypothesis #3. This finding is well consistent with previous studies that have proposed the reduced integration and segregation of information within the large-scale brain networks in ASD[43].

Furthermore, as illustrated in Supplementary Fig. S2, different regions from various Yeo sub-networks are present in high energy $T_1$ and $T_0$ triads in the CON network yet absent in the ASD network, among which regions from three sub-networks are dominant: the DMN, the SN, and the central executive network (CEN). Most involved regions in high energy $T_1$ and $T_0$ triads from the SN are the insula, the frontal opercular (FrOper) and the parietal opercular (ParOper), from the DMN the prefrontal cortex (PFC), the precuneus and posterior cingulate cortex (PCUN/PCC) as well as the inferior parietal lobule (IPL), and from the CEN mostly regions from the intraparietal sulcus (IPS). Interaction between these three sub-networks, known as a triple network model of the brain, has been recently proposed to underlie a wide range of disorders, including ASD[45]. *Zero participation of important regions of the DMN, the SN and the CEN in high energy $T_1$ and $T_0$ triads in the ASD network compered to the CON network, provides evidence for the triple network model from a perspective of SBT.* That is to say, high energy $T_1$ and $T_0$ triads provide a needed structure for a consistent and reliable three-way interaction between these three sub-networks which seems to be missing in ASD group.

*Narrow scale of change in energy of the whole brain network in ASD during development* (Fig. 3A). Our results have revealed that from 1st childhood to adolescence, there is an overall increase in energy of the whole-brain networks in both ASD and CON groups. We can thus hypothesize that increase of the whole-brain energy with age provides networks with the necessary structure to accommodate more dynamism needed for higher cognitive abilities, which are also increasing with age. Yet as results have revealed, the pattern of this increase from 1st childhood to 2nd childhood and from there to adolescence is significantly different between ASD and CON groups. That is, while networks in CON group start from being more balance during 1st childhood and gradually gain more energy during development, networks in ASD group start with higher mean energy during 1st childhood and then seem to be frozen after 2nd childhood. In other words, in ASD group the increasing change in energy of the whole-brain network is missed from 2nd childhood to adolescence, while it is significant in CON group.

It is worth mentioning that the whole-brain ASD network during 1st childhood having significantly higher energy than the CON network is due to the ASD network having practically less $T_1$ triads while the frequency of unbalanced triads is nearly the same in both networks (Fig. 1B). As both groups have nearly equal number of unbalanced triads, the higher energy of the ASD network during 1st childhood cannot be assigned to more dynamism, but to the less functional integration and segregation. This is because, as discussed above, $T_1$ triads provide needed structure for the integration between local segregated modules in networks. The severity and total ASD symptoms, as measured with ADOS-G, seem to confirm these results as well. That is, less energy during 1st childhood is associated with the severity of ASD. Altogether, our results have shown that *changes in the energy of the whole brain ASD networks during development are limited to a narrower band compared to CON group*—hypothesis #4.1. Moreover, during 2nd childhood and adolescence, although the total energy levels are not different between the ASD and CON networks, the underlying triadic interactions that give birth to these final energies are different (as has been discussed previously). Finally, during 2nd childhood the whole brain energy, as measured in the context of SBT, is positively associated with ADOS-G social scores and negatively with restricted and repetitive behaviors. While during adolescence, the whole-brain energy is positively correlated with restricted and repetitive behaviors as measured using ADI.

*Less energy of the SN and the DMN in ASD and its association with stereotype or repetitive behaviors* (Fig. 3B,C). We have observed that generally during development, the SN (A) and the DMN (B) are more balanced, have less energy, in the ASD than CON networks. In other words, in these two sub-networks in ASD group unbalanced triads are in minority compared to balanced triads. This is while unbalanced triads are known to be sources of dynamism through injecting energy into networks. That is to say, unbalanced triads are structures that excite a network towards change, unlike balanced triads that drive a network back to its stable states, that is, the minimum level of energy. Thus, balanced triads although underrepresented compared to chance, are playing a crucial role in healthy networks. As Heider himself stated, in a healthy community there may be a tendency to leave the balanced comfortable equilibrium to seek new horizons[17]. In other words, unbalanced triads are necessary if a community is to leave its comfort zone towards new experiments, that is, growth. As our results indicate, the mean energy of the SN (A) is nearly practically greater in the CON network compared to the ASD network during 1st childhood. The SN (A) in the Schaefer atlas[46] includes eight regions from the insula (Fig. 3D), namely, insula left 1 (−38, 2, −4), insula left 2 (−40, −14, −2), insula left 3 (−32, 18, 8), insula left 4 (−36, 4, 10), insula right 1 (40, 6, −16), insula right 2 (40, 8, −2), insula right 3 (40, −10, −4) and insula right 4 (40, −2, 6). Classically, the insula has been considered a limbic region, yet recent network neuroscience studies have suggested its vital role in detecting novel salient stimuli across multiple modalities. The SN itself is known to be involved in attentional processes and dynamic switching between task-positive (CEN) and task-negative (DMN) processes[47]. Moreover, the ability to detect and attend from one event to the other flexibly has been associated with the SN's well functioning. It is noteworthy that our finding shows a negative association between energy of the SN (A) and stereotype behaviors during 1st childhood. To be specific, greater energy of the SN (A) is associated with reduced stereotype behaviors during 1st childhood. Altogether, it seems that the *SN (A) having less energy in the ASD network is reflecting the difficulty in dynamic switching, which is manifested in form of increased repetitive and restricted behaviors*—hypothesis #4.2.

Furthermore, our results have unveiled that energy of the DMN (B) is practically greater in CON group than in ASD group during 2nd childhood. The DMN (B) in the Schaefer atlas[46] heavily relies on the regions from the prefrontal cortex (PFC) on both hemispheres, such as the ventral PFC, the dorsal PFC as well as the lateral PFC (Fig. 3E). The role of the PFC on both social impairment and restricted behaviors in ASD has been suggested, specifically the proper level of dopamine in the PFC seems to be necessary for jumping out of repetitive behaviors[48]. For example, when an antagonist of dopamine was injected into the PFC of rats, it induced ipsiversive circling[49]. From a network point of view, it is known that systems with more energy are more probable to explore different possible states, while for networks with less energy the chance to be trapped in one of the local minima is higher. *Less energy of the DMN (B), which includes several regions from the PFC, seems to be a candidate explanation for a deficiency to move from one local minimum to another in the ASD network, which may be expressing itself as increased repetitive behaviors*—**hypothesis #4.3**. The general pattern of association between energy and restricted behaviors, the more the energy of the DMN (B) the less repetitive behaviors, nearly holds for the DMN (B) during 2nd childhood as well. Although it did not reach significance ($p = 0.10$, $r = -0.23$) due to missing values, thus further investigations can be helpful in this regard.

All things considered, the current study proposes SBT as a promising context to understand alterations found in the brain networks of individuals with ASD compared to healthy controls. A limitation of this study is that the data analyzed here are cross-sectional due to the small sample size of open longitudinal fMRI dataset. Thus, further investigations based on longitudinal data can sure be insightful. According to the results obtained here, studying triadic interactions and the consequent structural balance of the weighted signed brain networks, and the role that their disruption may play on the complex neurodevelopmental disorders seems to be of value. Furthermore, in addition to SBT for studying higher-order interactions in the brain networks methods such as the topological data analysis can be insightful.

## Methods

**Participants.** T1-weighted MRI and resting-state fMRI along with the corresponding demographic data of 311 individuals, 152 with ASD, and 159 healthy volunteers (CON), aggregated from multiple sites in the publicly available Autism Brain Imaging Data Exchange (ABIDE I Preprocessed repository)[50], were processed and analyzed in this study (Fig. 5A). Inclusion criteria for sites were to have (1) Similar repetition time ($TR = 2\ ms$), in order to limit the multi-site variability and (2) Instructed participants to relax with their eyes open while a cross was projected on a screen. The reason for this choice was that, as has been reported previously[51], the reliability of resting-state analysis is higher when subjects are instructed to rest with their eyes open compared to the eyes closed, due to the possibility of falling sleep when eyes closed. These two criteria resulted in five sites, namely, New York University Langone Medical Center (NYU), San Diego State University (SDSU), University of Michigan (both samples: UM-1, UM-2), University of Utah School of Medicine (USM) and Yale Child Study Center (Yale). Furthermore, it is worth mentioning that in the NYU dataset 14% of subjects were eyes closed, which have been excluded prior to preprocessing.

Inclusion as a participant in ASD group required receiving ASD diagnosis based on the autism diagnostic observation schedule-generic (ADOS-G) and an expert clinical opinion upon DSM-IV criteria. Individuals in CON group must have no history of psychiatric disorders in themselves or their first-degree relatives. Participants in both groups should have no prior or concurrent diagnosis of neurological disorders (e.g., epilepsy, meningitis, encephalitis, head trauma, or seizures). Being fully verbal and IQ > 70, as has been measured via WASI and/ or WASI-IV, were also required for all. Moreover, to met the assumptions of ANCOVA we have discarded 53 participants due to being outliers (Fig. 5D), details of which are provided in statistical analysis. Altogether, the final number of participants has been 258 individuals, 128 with ASD and 130 healthy controls. We have further divided these individuals based on age, that is, 6–13 years old as children and 13–18 years old as adolescents. Additionally, as middle childhood is a crucial period during development, we have considered the opening years of middle childhood and the closing years separately. We have referred to the former as 1st childhood (6–9 years old) and the latter as 2nd childhood (9–13 years old), that are age, gender, handedness, IQ and head motion matched (Table 3).

**Approval for human experiments.** For each of the included sites following Institutional Review Boards (IRB) and ethics committees have approved the experiments, and data acquisition procedures were in accordance to their guidelines and regulations: (1) IRB Operations at NYU, (2) San Diego State University's Human Research Protection Program (3) IRB of the University of Michigan Medical School (4) The University of Utah's IRB (5) Yale University's IRB. Furthermore, all experiments were in accordance with HIPAA guidelines and 1000 Functional Connectomics Project / INDI protocols, that is, all data were fully anonymized with no protected health information and legal guardians of all participants have signed informed consent.

**Data preprocessing.** We have used the preprocessed version of ABIDE I neuro-imaging data that is provided by the Preprocessed Connectomes Project (PCP) and is public at https://preprocessed-connectomes-project.org/abide/. Among the four available preprocessing pipelines in ABIDE I preprocessed, we have used data from Configurable Pipeline for the Analysis of Connectomes (CPAC). The following preprocessing steps were performed: (1) Basic processing that includes: Realignment, slice timing correction, registration (Boundary-based rigid body (BBR) for functional to anatomical and ANTs for anatomical to standard template), intensity normalization (4D Global mean = 1000). (2) Nuisance signal removal to clean confounding variations resulting from head motion, physiological processes (such as heart beat and respiration) and scanner drifts that have included regressing: 24 head motion parameters (six standard parameters $R = [\ X\ Y\ Z$ pitch yaw roll], along with their squares $R^2$ and temporal derivatives $R'$), tissue signals using component-based noise correction method (CompCor[52]), linear and quadratic scanner's low frequency drifts. (3) As it has been reported[53], even after regressing head motion parameters resting-state functional connectivity may still be affected by motion thus scans with frame-wise displacement > 0.5 mm and global BOLD signal changes > 3 SD (the conservative option in the default preprocessing steps of Conn toolbox[52]) were flagged and scrubbed. Participants with at least 150 valid scans ($\sim$ 5 min or more) have been included. Moreover, results of the Mann-Whitney U test have confirmed that the difference between ASD and CON groups was not statistically significant regarding head motion as has been quantified by mean frame-wise displacement (Table 3). (4) We have not regressed the global signal out, that is, the global mean signal was not included with nuisance variable regression. Moreover, spatial smoothing using a Gaussian kernel of 6 mm FWHM has been conducted. We have visually inspected the de-noised data using Conn QA plots[52]. Band-pass filtering (Slow-4: 0.027–0.073 HZ) was applied after preprocessing and during the de-noising step because it has been suggested that resting-state brain networks derived in this frequency exhibit grater reliability[54]. (5) Finally, to harmonized data across the five sites MatLab implementation of the ComBat method[55] have been applied through following steps: (1) For each $i = 1$ to 258 subjects we have created the $tril_i$ (79,800 $\times$ 1) matrix with the lower triangle elements of its (400 $\times$ 400) functional connectivity matrix. Then, we have designed the *dat* input matrix by concatenating *tril* matrices column-wise, resulting in a (79,800 $\times$ 258) matrix with columns regarding subjects and rows corresponding to their functional connectivities. (2) We have coded different sites from 1 to 5 corresponding to NYU, SDSU, UM, USM and YALE, respectively. Then, we have set the *batch* input to be a (258 $\times$ 1) matrix, whose element $(i, 1)$ corresponds to subjects $i$ site code. (3) For the *mod* input, which aims to preserve the original biological variations within data while harmonizing the effect of site, we have designed a (258 $\times$ 2) matrix with first and second columns being age and group (0 for ASD, 1 for CON), respectively. (4) We have conducted a series of Kruskal–Wallis tests (FDR corrected $p$ values at 5% significance) to investigate site effect on functional connectivities between all pair regions of interest before and after applying ComBat harmonization. As results have shown before the harmonization 1.37% of all functional connectivities were significantly different across the five sites, which have decreased to zero after harmonization.

| | 1st Childhood, (6–9) years | | | 2nd Childhood, (9–13) years | | | Adolescence, (13–18) years | | |
|---|---|---|---|---|---|---|---|---|---|
| Group size | CON (n = 20) | ASD (n = 18) | p val | CON (n = 52) | ASD (n = 52) | p val | CON (n = 58) | ASD (n = 58) | p val |
| Age | 8.21 ± 0.68 | 8.08 ± 0.74 | 0.63 | 11.14 ± 1.06 | 11.02 ± 1.14 | 0.61 | 16.04 ± 1.45 | 15.93 ± 1.56 | 0.80 |
| Gender, male | 15 | 18 | 0.13 | All male | All male | 1 | All male | All male | 1 |
| Hand, right | 19 | 15 | 0.21 | 47 | 42 | 0.16 | 54 | 50 | 0.25 |
| Full IQ | 115.90 ± 13.08 | 112.30 ± 20.53 | 0.50 | 111.04 ± 14.20 | 103.86 ± 18.85 | 0.03* | 108.89 ± 11.87 | 102.82 ± 15.88 | 0.02* |
| Verbal IQ | 114.14 ± 14.83 | 107.50 ± 17.23 | 0.19 | 113.67 ± 12.50 | 105.44 ± 20.63 | 0.01* | 111.34 ± 13.31 | 99.69 ± 20.21 | < 0.001** |
| Perform IQ | 112.95 ± 13.70 | 114.05 ± 23.98 | 0.85 | 105.98 ± 16.38 | 102.50 ± 19.78 | 0.33 | 105.10 ± 12.85 | 105.69 ± 15.96 | 0.82 |
| Mean FD | 0.14 ± 0.05 | 0.19 ± 0.06 | 0.06 | 0.19 ± 0.09 | 0.20 ± 0.09 | 0.70 | 0.14 ± 0.05 | 0.17 ± 0.08 | 0.10 |
| ADOS-G Total | – | 11.95 ± 4.67 | – | – | 12.11 ± 5.49 | – | – | 12.87 ± 4.98 | – |
| ADOS-G Social | – | 8.95 ± 3.64 | – | – | 8.92 ± 4.27 | – | – | 10.10 ± 4.21 | – |
| ADOS-G RRB | – | 3.45 ± 1.50 | – | – | 3.25 ± 1.76 | – | – | 3.00 ± 1.77 | – |
| ADI-R Verbal | – | 16.75 ± 5.10 | – | – | 15.77 ± 4.00 | – | – | 16.35 ± 3.55 | – |
| ADI-R Social | – | 19.00 ± 6.06 | – | – | 19.39 ± 5.27 | – | – | 19.82 ± 5.28 | – |
| ADI-R RRB | – | 5.75 ± 3.10 | – | – | 6.36 ± 2.81 | – | – | 6.32 ± 2.33 | – |

**Table 3.** Demographic data. $p$ values are according to paired t-test for full/verbal/perform IQ, Mann–Whitney U test for age, mean FD and Chi-square test for gender and handedness. ** $p < 0.01$, * $p < 0.05$. Mean ± std, ADOS-G, the autism diagnostic observation schedule-generic; ADI-R, the autism diagnostic interview-revised; FD, frame-wise displacement; RRB, restricted and repetitive behaviors; CON, control; ASD, autism spectrum disorder.

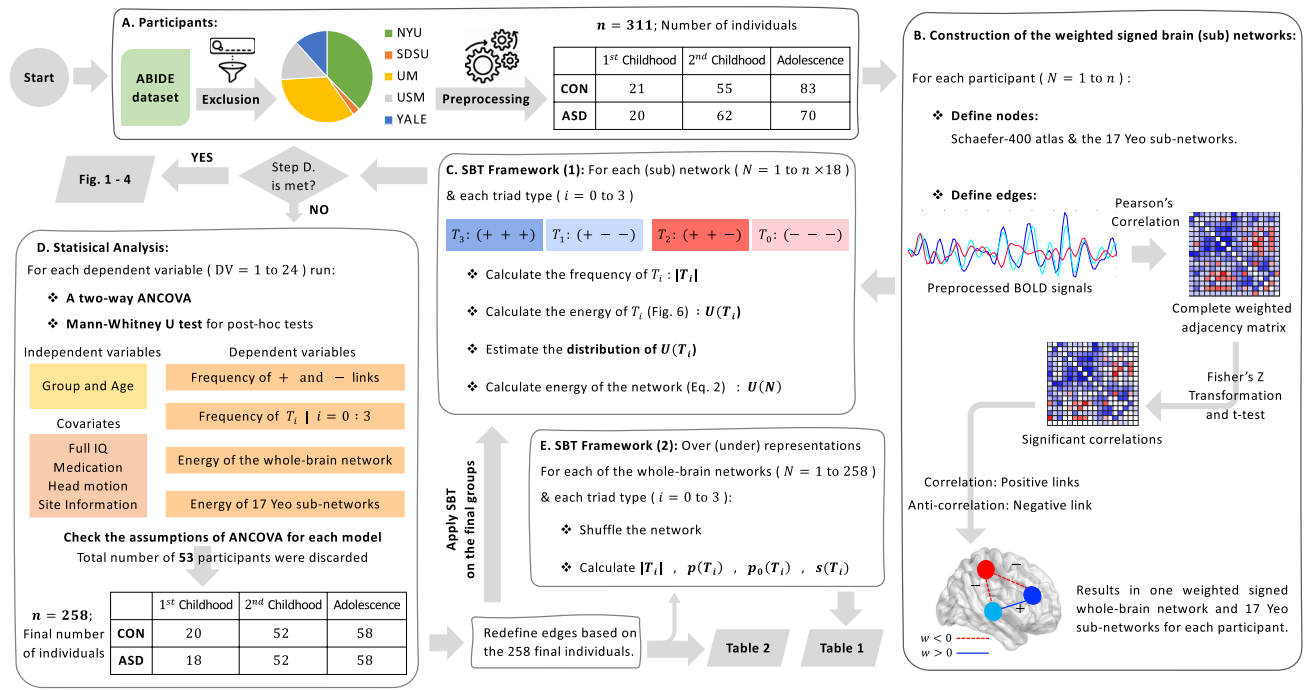

**Figure 5.** *Graphical abstract* (**A**) After preprocessing, 311 individuals (159 and 152 in CON and ASD groups, respectively) have been included. (**B**) Brain construction was based on Schaefer-400 atlas as nodes and significant Pearson's correlations as edges. (**C,E**) Procedures as has been defined in the context of SBT. **D** Statistical analysis for comparing group means. For head motion and site information covariates, frame-wise displacement and site codes have been used, respectively. As a results of applying ANCOVA's assumption 53 individuals were discarded. After excluding these individuals, network's edges have been redefined based on the remaining 258 participants. Step **C** has been reapplied on the final networks. BrainNet Viewer 1.63[40] (nitrc.org/projects/bnv) has been used for visualization of the brain networks. n, number of individuals;$|T_i|$, total number of $T_i$; $p(T_i)$, fraction of $T_i$; $p_0(T_i)$, fraction of $T_i$ in the null model; $s(T_i)$, the amount of surprise; SBT, structural balance theory; NYU, New York University; SDSU, San Diego State University; UM, University of Michigan; USM, University of Utah School of Medicine; YALE, Yale Child Study Center; CON, Control; ASD, autism spectrum disorder.

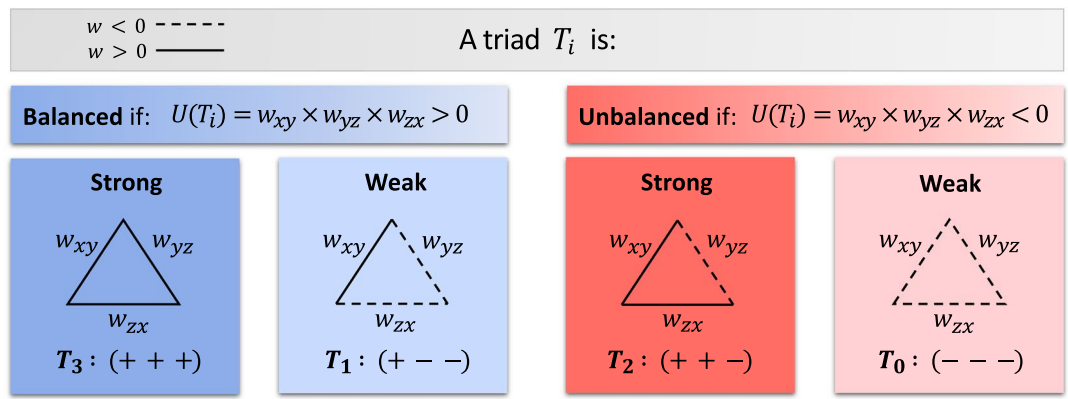

**Figure 6.** Four types of triads as has been defined in SBT. The number of negative links is even for balanced triads, and odd for unbalanced triads. Furthermore, being strong or weak refers to how much frustration a presence of a triad imposes on a network, with $T_2$ triads injecting more frustration than $T_0$ triads, and $T_3$ being more stable than $T_1$ triads[14]. $w$, weights of links; $U(T_i)$, energy of $T_i$.

**Construction of the brain networks from resting-state fMRI data.** In this study, we have defined the whole brain networks' nodes based on 400 regions of interest as introduced by the 2-mm Schaefer-400 atlas[46]. Moreover, for the analysis of brain sub-networks, 17 Yeo parcellation was applied as defined in the Schaefer-400 atlas as well. Specifically, on this atlas the DMN (B) and the SN (A) have 32 and 34 regions of interest, respectively. For links of the whole brain network, we have assessed functional connectivity using Pearson's correlation between all pairs of 400 regions of interest for each participant across the full length of the resting-state Blood Oxygen Level-Dependent (BOLD) time series, creating 311 weighted correlation matrices $r(400 \times 400)$. The same calculations have been carried for the 17 Yeo functional sub-networks. Next, as Fisher's z-transformation stabilizes the variance for better use in statistical testings, we have converted Pearson's r to the normally distributed z variables. Then, we have performed two-tailed one sample t-tests on Fisher's Z-transformed correlation coefficients (z), to check whether correlation coefficients are spurious or significantly different from zero[1]. Specifically, for each edge $(i, j)$ we have performed a one-sample t-test based on the distribution of z values between $i$ and $j$ across the group. Then, for every $(i, j)$ we have set it to zero if it did not pass the 5% significance. Of course, a multiple comparison correction was necessary to account for multiple comparisons, thus Bonferroni corrected $p$ values were used. We applied no further manual threshold on networks to avoid its inevitable effects on network parameters. This procedure has been conducted on MatLab and Statistics Toolbox Release 2017b and the brain networks have been visualized in BrainNet Viewer version 1.63[40] (Fig. 5B).

**Structural balance theory.** Our approach to studying the brain networks is structural balance theory (SBT) that provides a framework to go beyond the assumption that pair interactions are independent from each other in signed networks, through analyzing triadic interactions[18,32]. Similar to the graph-theoretical framework, SBT has also been developed to investigate the organizational properties of complex networks[31]. However, unlike graph theory, which has been formulated initially to model dyadic relations between information units, SBT goes beyond pairwise interactions and study triadic interactions. Specifically, SBT proposes that a network evolves in a direction that leads to the minimum level of tension and frustration between triadic interrelationships[56,57]. When applying graph theoretical analysis on the brain networks, one defines a graph $G(V, E)$ with a set of nodes $V = \{v_n; \ n = 1, 2, \dots, N\}$ and estimates a measure of association between each pair regions of interest as a set of weighted links $E = \{w_{xy} \mid w_{xy} > 0 \text{ if } v_x \text{ is correlated with } v_y, \text{ and } w_{xy} < 0 \text{ if } v_x \text{ is anti-correlated with } v_y\}$. Yet, in the context of SBT as network's primary building blocks one defines sets of triads as follows[14] (Fig. 6):

$$T_i = \{\Delta_{xyz} \mid v_x, \ v_y \text{ and } \ v_z \text{ are (anti) associated}\}, \text{where } i = \begin{cases} 3: \text{ strongly balanced,} & \text{if } w_{xy}, w_{yz}, w_{zx} > 0 \\ 2: \text{ strongly unbalanced,} & \text{if } w_{xy}, w_{yz} > 0 \text{ and } w_{zx} < 0 \\ 1: \text{ weakly balanced,} & \text{if } w_{xy}, w_{yz} < 0 \text{ and } w_{zx} > 0 \\ 0: \text{ weakly unbalanced,} & \text{if } w_{xy}, w_{yz}, w_{zx} < 0 \end{cases}$$

$$(1)$$

This way, information about the organization of network that cannot be detected on the level of pair connections would get a chance to be revealed. A well-known analogy for this definition is that positive (negative) links are considered friendship (enmity) relations, respectively. Then, balanced triads are those that fulfill following axioms: (1) A friend of my friend is my friend, that is, strongly balanced triad, $T_3$ (2) An enemy of my enemy is my friend, a friend of my enemy is my enemy and an enemy of my friend is my enemy, that is, weakly balanced triads, $T_1$, otherwise we have unbalanced triads, $T_2$ and $T_0$. A network satisfies the structural balance property in two ways: (1) Either all its triads are balanced (known as heaven), or (2) It is divided into sub-networks such that within each sub-network positive links are present, while there are negative links between sub-networks (known as bipolar). In this study, we have taken the correlations (anti-correlations) between each two regions of interest in the brain network as positive (negative) relations, respectively.

It can be seen from Eq. 1 that, for a balanced interaction the product of its edges is a positive number, whereas for an unbalanced interaction this product is negative. As has been previously proposed[58], if one sums the negative of these products and divides it by the total number of ternary interactions, a quantity $U(N)$ or energy of a network would be obtained. Energy represents the extent to which a network is structurally balanced. Explicitly,

$$U(N) = -\frac{1}{\Delta} \sum_{x < y < z} w_{xy} w_{yz} w_{zx}$$

$$(2)$$

in which $N$ is the number of nodes and $\Delta$ is the total number of triads in the network. Applying this measure on a fully balanced network results in a lowest $U(N)$, that is, $-1$, while a fully unbalanced network obtains the highest possible $U(N)$, that is, $+1$. In this study, for any given network we have:

- Counted the number (frequency) of $T_i$ ($|T_i|$), then calculated the energy of $T_i$, i.e., $U(T_i) = w_{xy} w_{yz} w_{zx}$, where $x$, $y$ and $z$ are $T_i$'s nodes and $w_{xy}$, $w_{yz}$ and $w_{zx}$ are link's weights. Afterwards, we have estimated the distribution of $U(T_i)$ during 1nd childhood, 2nd childhood and adolescence for both ASD and CON groups. Finally, we have computed total energy of the network, $U(N)$ (Fig. 5C).
- As has been previously proposed[12], we have created a null model, which is a network with the exact fraction of positive (negative) signs which are randomly assigned to existing links. The purpose of this null model is to compare the empirical frequencies of $T_i$, as in the real brain networks, with the corresponding frequencies if signs of links were generated randomly from the same distribution of positive (negative) signs. In other words, the null model represents a condition where no underlying structure (organization) directs the placement of signs, rather it is random. Thus, after generating this shuffled version of a given network we have computed $p_0(T_i)$, fraction of triad $T_i$ in the null model, and compared it to $p(T_i)$, fraction of triad $T_i$ in the real brain network. If $p(T_i) > p_0(T_i)$ then $T_i$ is overrepresented, and if $p(T_i) < p_0(T_i)$ then $T_i$ is underrepresented. Furthermore, we have calculated the value of surprise, $s(T_i) = (T_i - E[T_i])/\sqrt{\Delta p_0(T_i)(1 - p_0(T_i))}$, in which $E[T_i] = p_0(T_i)\Delta$ is the expected number of triads $T_i$ and $\Delta$ is the total number of triads in the network (Fig. 5E).

**Statistical analysis.** Throughout this study, to determine the group mean differences we have conducted two-way Analysis of Covariance (ANCOVA) with group and age as independent variables while controlling for FIQ, medication, mean frame-wise displacement as head motion parameter and site information. In total, we had 24 dependent variables, namely, the frequency of positive and negative links (Table 2A), the frequency of each types of triads (Table 2B), energy of the whole-brain network (Fig. 3A), energy of each of the 17 Yeo sub-networks among which results of ANCOVA were significant for SN (A) and DMN (B) (Fig. 3B,C). For each of these 24 models, first we have checked the assumptions of ANCOVA regarding the dependent variable of that model as follows: (1) We have explored if the dependent variable have outliers across different groups of independent variables using box plots, which in total have resulted in discarding 53 out of 311 participants. (2) Results of the Kolmogorov–Smirnov tests have indicated that except for the frequency of $T_2$ triads in ASD group during 2nd childhood and adolescence, as well as the frequency of $T_0$ triads in ASD group during adolescence, which are only moderately skewed ($-1 \leq skewness \leq +1$), all other dependent variables in this study are normally distributed. (3) Homogeneity of variances was investigated through Levene's Test of Equality of Error Variances. (4) The linear relation between the dependent variable and covariates was studied through scatter matrices. Additionally, when results of ANCOVA were significant we have applied the Mann-Whitney U test as the post-hoc test to determine the specific groups that are different (as shown using the asterisks above the box plots in Fig. 1 for the frequency of links and triads as dependent variable, and Fig. 3A–C, for energy of the whole brain and sub-networks as dependent variables). In case of ANCOVA, effect sizes are reported as has been readily estimated by IBM SPSS Statistics version 26, that is, $partial\eta^2 = SS_{effect}/SS_{effect} + SS_{error}$. For the interpretation, Cohen's guideline was applied, i.e., $\eta^2$ at least 0.01: small, 0.06: medium and 0.14: large effects[59]. Likewise, for the post-hoc tests we have calculated $\eta^2 = Z^2/n - 1$, using the $Z$ statistics from the Mann-Whitney U test.

Finally, the Kullback–Leiber Divergence between distributions $P$ and $Q$ in Fig. 2 has been calculated through $\mathbb{K}(P||Q) + \mathbb{K}(Q||P) = \sum_i \log_2(p_i/q_i)p_i + \sum_i \log_2(q_i/p_i)q_i$.

## Data availability

The datasets analysed during the current study are available from the corresponding author upon request.

## References

1. Fornito, A., Zalesky, A. & Bullmore, E. *Fundamentals of Brain Network Analysis* (Academic Press, Cambridge, 2016).
2. Sporns, O. *Networks of the Brain* (MIT Press, Cambridge, 2010).
3. Watts, D. J. & Strogatz, S. H. Collective dynamics of 'small-world' networks. *Nature* **393**, 440–442. https://doi.org/10.1038/30918 (1998).
4. Sporns, O., Chialvo, D. R., Kaiser, M. & Hilgetag, C. C. Organization, development and function of complex brain networks. *Trends Cognit. Sci.* **8**, 418–425. https://doi.org/10.1016/j.tics.2004.07.008 (2004).
5. Bullmore, E. & Sporns, O. Complex brain networks: graph theoretical analysis of structural and functional systems. *Nat. Rev. Neurosci.* **10**, 186–198. https://doi.org/10.1038/nrn2575 (2009).
6. Long, Z., Duan, X., Mantini, D. & Chen, H. Alteration of functional connectivity in autism spectrum disorder: effect of age and anatomical distance. *Sci. Rep.* **6**, 26527. https://doi.org/10.1038/srep26527 (2016).
7. Sadeghi, M. *et al.* Screening of autism based on task-free fmri using graph theoretical approach. *Psychiatry Res. Neuroimaging* **263**, 48–56. https://doi.org/10.1016/j.pscychresns.2017.02.004 (2017).
8. Lau, W. K., Leung, M.-K. & Lau, B. W. Resting-state abnormalities in autism spectrum disorders: a meta-analysis. *Sci. Rep.* **9**, 1–8. https://doi.org/10.1038/s41598-019-40427-7 (2019).
9. He, Y., Zhou, Y., Ma, W. & Wang, J. An integrated transcriptomic analysis of autism spectrum disorder. *Sci. Rep.* **9**, 1–9. https://doi.org/10.1038/s41598-019-48160-x (2019).
10. Lord, C. *et al.* Autism spectrum disorder. *Nat. Rev. Dis. Primers* **6**, 5. https://doi.org/10.1038/s41572-019-0138-4 (2020).
11. Sherkatghanad, Z. *et al.* Automated detection of autism spectrum disorder using a convolutional neural network. *Front. Neurosci.* https://doi.org/10.3389/fnins.2019.01325 (2019).
12. Leskovec, J., Huttenlocher, D. & Kleinberg, J. Signed networks in social media. *Proc. SIGCHI Conf. Hum. Factors Comput. Syst.* **22**, 1361–1370. https://doi.org/10.1038/309180 (2010).
13. Szell, M., Lambiotte, R. & Thurner, S. Multirelational organization of large-scale social networks in an online world. *Proc. Natl. Acad. Sci.* **107**, 13636–13641. https://doi.org/10.1038/309181 (2010).
14. Belaza, A. M. *et al.* Statistical physics of balance theory. *PLoS ONE* **12**, e0183696. https://doi.org/10.1038/309182 (2017).
15. Doreian, P. & Mrvar, A. Structural balance and signed international relations. *J. Soc. Struct.* **16**, 1. https://doi.org/10.1038/309183 (2015).
16. Saiz, H. *et al.* Evidence of structural balance in spatial ecological networks. *Ecography* **40**, 733–741. https://doi.org/10.1038/309184 (2017).
17. Heider, F. *The Psychology of Interpersonal Relations* (Psychology Press, Hove, 1982).
18. Cartwright, D. & Harary, F. Structural balance: a generalization of heider's theory. *Psychol. Rev.* **63**, 277. https://doi.org/10.1038/309185 (1956).
19. Rizi, A.K., Zamani, M., Shirazi, A., Jafari, G.R. & Kertész, J. Stability of imbalanced triangles in gene regulatory networks of cancerous and normal cells. arXiv preprint https://doi.org/10.1038/309186 (2020).
20. Sporns, O. & Kötter, R. Motifs in brain networks. *PLoS Biol.* **2**, e369. https://doi.org/10.1038/309187 (2004).
21. Alon, U. Network motifs: theory and experimental approaches. *Nat. Rev. Gen.* **8**, 450–461. https://doi.org/10.1038/309188 (2007).
22. Rubinov, M. & Sporns, O. Complex network measures of brain connectivity: uses and interpretations. *Neuroimage* **52**, 1059–1069. https://doi.org/10.1038/309189 (2010).
23. Borgatti, S. P., Mehra, A., Brass, D. J. & Labianca, G. Network analysis in the social sciences. *Science* **323**, 892–895. https://doi.org/10.1016/j.tics.2004.07.080 (2009).
24. Facchetti, G., Iacono, G. & Altafini, C. Computing global structural balance in large-scale signed social networks. *Proc. Natl. Acad. Sci.* **108**, 20953–20958. https://doi.org/10.1016/j.tics.2004.07.0081 (2011).
25. Kirkley, A., Cantwell, G. T. & Newman, M. Balance in signed networks. *Phys. Rev. E* **99**, 012320. https://doi.org/10.1016/j.tics.2004.07.0082 (2019).
26. Sontag, E. D. Monotone and near-monotone biochemical networks. *Syst. Synth. Biol.* **1**, 59–87. https://doi.org/10.1016/j.tics.2004.07.0083 (2007).
27. Fox, M. D. *et al.* The human brain is intrinsically organized into dynamic, anticorrelated functional networks. *Proc. Natl. Acad. Sci.* **102**, 9673–9678. https://doi.org/10.1016/j.tics.2004.07.0084 (2005).
28. Liang, Z., King, J. & Zhang, N. Anticorrelated resting-state functional connectivity in awake rat brain. *Neuroimage* **59**, 1190–1199. https://doi.org/10.1016/j.tics.2004.07.0085 (2012).
29. Fox, M. D., Zhang, D., Snyder, A. Z. & Raichle, M. E. The global signal and observed anticorrelated resting state brain networks. *J. Neurophysiol.* **101**, 3270–3283. https://doi.org/10.1016/j.tics.2004.07.0086 (2009).
30. Murphy, K., Birn, R. M., Handwerker, D. A., Jones, T. B. & Bandettini, P. A. The impact of global signal regression on resting state correlations: are anti-correlated networks introduced?. *Neuroimage* **44**, 893–905. https://doi.org/10.1016/j.tics.2004.07.0087 (2009).
31. Antal, T., Krapivsky, P. L. & Redner, S. Social balance on networks: the dynamics of friendship and enmity. *Physica D* **224**, 130–136. https://doi.org/10.1016/j.tics.2004.07.0088 (2006).
32. Kułakowski, K., Gawroński, P. & Gronek, P. The heider balance: a continuous approach. *Int. J. Mod. Phys. C* **16**, 707–716. https://doi.org/10.1016/j.tics.2004.07.0089 (2005).
33. Krawczyk, M. J., Wołoszyn, M., Gronek, P., Kułakowski, K. & Mucha, J. The heider balance and the looking-glass self: modelling dynamics of social relations. *Sci. Rep.* **9**, 1–8. https://doi.org/10.1038/nrn25759 (2019).
34. Górski, P. J., Kułakowski, K., Gawroński, P. & Hołyst, J. A. Destructive influence of interlayer coupling on heider balance in bilayer networks. *Sci. Rep.* **7**, 1–12. https://doi.org/10.1038/nrn25751 (2017).
35. Kargaran, A., Ebrahimi, M., Riazi, M., Hosseiny, A. & Jafari, G. R. Quartic balance theory: global minimum with imbalanced triangles. *Phys. Rev. E* **102**, 012310. https://doi.org/10.1038/nrn25752 (2020).
36. Chiang, Y.-S., Chen, Y.-W., Chuang, W.-C., Wu, C.-I. & Wu, C.-T. Triadic balance in the brain: seeking brain evidence for Heider's structural balance theory. *Soc. Netw.* **63**, 80–90. https://doi.org/10.1038/nrn25753 (2020).

37. Tadić, B., Andjelković, M., Boshkoska, B. M. & Levnajić, Z. Algebraic topology of multi-brain connectivity networks reveals dissimilarity in functional patterns during spoken communications. *PLoS ONE* **11**, e0166787. https://doi.org/10.1038/nrn2575 4 (2016).

38. Tadić, B., Andjelković, M. & Melnik, R. Functional geometry of human connectomes. *Sci. Rep.* **9**, 1–12. https://doi.org/10.1038/nrn25755 (2019).

39. Zhu, H. *et al.* Altered topological properties of brain networks in social anxiety disorder: a resting-state functional MRI study. *Sci. Rep.* **7**, 43089. https://doi.org/10.1038/nrn25756 (2017).

40. Xia, M., Wang, J. & He, Y. Brainnet viewer: a network visualization tool for human brain connectomics. *Plos ONE* https://doi.org/10.1371/journal.pone.0068910 (2013).

41. Supekar, K. *et al.* Brain hyperconnectivity in children with autism and its links to social deficits. *Cell Rep.* **5**, 738–747. https://doi.org/10.1016/j.celrep.2013.10.001 (2013).

42. Uddin, L. Q., Supekar, K. & Menon, V. Reconceptualizing functional brain connectivity in autism from a developmental perspective. *Front. Hum. Neurosci.* **7**, 458. https://doi.org/10.3389/fnhum.2013.00458 (2013).

43. Rudie, J. D. *et al.* Altered functional and structural brain network organization in autism. *Neuroimage Clin.* **2**, 79–94. https://doi.org/10.1016/j.nicl.2012.11.006 (2013).

44. Peper, J. S., van den Heuvel, M. P., Mandl, R. C., Pol, H. E. H. & van Honk, J. Sex steroids and connectivity in the human brain: a review of neuroimaging studies. *Psychoneuroendocrinology* **36**, 1101–1113. https://doi.org/10.1016/j.psyneuen.2011.05.004 (2011).

45. Menon, V. The triple network model, insight, and large-scale brain organization in autism. *Biol. Psychiatry* **84**, 236. https://doi.org/10.1016/j.biopsych.2018.06.012 (2018).

46. Schaefer, A. *et al.* Local-global parcellation of the human cerebral cortex from intrinsic functional connectivity MRI. *Cereb. Cortex* **28**, 3095–3114. https://doi.org/10.1093/cercor/bhx179 (2018).

47. Menon, V. & Uddin, L. Q. Saliency, switching, attention and control: a network model of insula function. *Brain Struct. Funct.* **214**, 655–667. https://doi.org/10.1007/s00429-010-0262-0 (2010).

48. Kim, H., Lim, C.-S. & Kaang, B.-K. Neuronal mechanisms and circuits underlying repetitive behaviors in mouse models of autism spectrum disorder. *Behav. Brain Funct.* **12**, 3. https://doi.org/10.1186/s12993-016-0087-y (2016).

49. Morency, M. A., Stewart, R. J. & Beninger, R. J. Effects of unilateral microinjections of sulpiride into the medial prefrontal cortex on circling behavior of rats. *Prog. Neuropsychopharmacol. Biol. Psychiatry* **9**, 735–738. https://doi.org/10.1016/0278-5846(85)90051-X (1985).

50. Craddock, C. *et al.* The neuro bureau preprocessing initiative: open sharing of preprocessed neuroimaging data and derivative

51. Noble, S., Scheinost, D. & Constable, R. T. A decade of test-retest reliability of functional connectivity: a systematic review and meta-analysis. *Neuroimage* **203**, 116157. https://doi.org/10.1016/j.neuroimage.2019.116157 (2019).

52. Whitfield-Gabrieli, S. & Nieto-Castanon, A. Conn: a functional connectivity toolbox for correlated and anticorrelated brain networks. *Brain Connect.* **2**, 125–141. https://doi.org/10.1089/brain.2012.0073 (2012).

53. Power, J. D. *et al.* Methods to detect, characterize, and remove motion artifact in resting state fMRI. *Neuroimage* **84**, 320–341. https://doi.org/10.1016/j.nicl.2012.11.0060 (2014).

54. Liang, X. *et al.* Effects of different correlation metrics and preprocessing factors on small-world brain functional networks: a resting-state functional MRI study. *Plos ONE* https://doi.org/10.1371/journal.pone.0032766 (2012).

55. Yu, M. *et al.* Statistical harmonization corrects site effects in functional connectivity measurements from multi-site fMRI data. *Hum. Brain Mapp.* **39**, 4213–4227. https://doi.org/10.1002/hbm.24241 (2018).

56. Rahbani, F., Shirazi, A. H. & Jafari, G. R. Mean-field solution of structural balance dynamics in nonzero temperature. *Phys. Rev. E* **99**, 062302. https://doi.org/10.1103/PhysRevE.99.062302 (2019).

57. Sheykhali, S., Darooneh, A. H. & Jafari, G. R. Partial balance in social networks with stubborn links. *Physica A Stat. Mech. Appl.* https://doi.org/10.1016/j.physa.2019.123882 *(2019).*

58. Marvel, S. A., Strogatz, S. H. & Kleinberg, J. M. Energy landscape of social balance. *Phys. Rev. Lett.* **103**, 198701. https://doi.org/10.1103/PhysRevLett.103.198701 (2009).

59. Cohen, J. *Statistical Power Analysis for the Behavioral Sciences* (Academic Press, Cambridge, 2013).

## Acknowledgements

We appreciate A. Kargaran, M. Saberi, M. Bagherikalhor and L. Noorbala for insightful discussions and helpful comments.

## Author contributions

Z.M., R.K., M.E.G., and G.R.J conceived and designed the study. Z.M., R.K., and G.R.J conducted the experiments. Z.M. analyzed the data, performed the statistical analysis, created the figures, and wrote the first draft of the manuscript. Z.M., R.K., M.E.G., and G.R.J analyzed the results. All authors read and approved the final manuscript.

## Competing interests

The authors declare no competing interests.

## Additional information

**Supplementary Information** The online version contains supplementary material available at https://doi.org/10.1038/s41598-020-80330-0.

**Correspondence** and requests for materials should be addressed to G.R.J.

**Reprints and permissions information** is available at www.nature.com/reprints.

**Publisher's note** Springer Nature remains neutral with regard to jurisdictional claims in published maps and institutional affiliations.